\documentstyle[aps,preprint,epsfig]{revtex}
\tighten
\begin{document}
\title{FROM KAONS TO NEUTRINOS: \\
QUANTUM MECHANICS OF PARTICLE OSCILLATIONS\thanks{
Presented at the XXXVIII Cracow School of Theoretical Physics, Zakopane,
June 1-10, 1998.}\hspace{1mm} \thanks{
 This work was supported by Polish Committee of Scientific Research under
Grants No. 2P03B08414 and 2P03B04215.}}

\author{Marek Zra\l ek  
\address{Department of Field Theory and Particle Physics\\ 
Institute of Physics,
University of Silesia\\ 
Uniwersytecka 4, 40-007 Katowice, Poland\\ e-mail:
zralek@us.edu.pl}}
\maketitle

\begin{abstract}
The problem of particle oscillation is considered in a pedagogical and
comprehensive way. Examples from K, B and neutrino physics are
given. 
Conceptual difficulties of the traditional approach to particle
oscillation are discussed. It is shown how the probability current density
and the wave packet treatments of particle oscillations resolve some
problems. It is also shown that only full field theoretical approach is free
from conceptual difficulties. The possibility of oscillation of particles
produced together with kaons or neutrinos is considered in full wave packet
quantum mechanics language. Precise definition of the oscillation of
particles which recoil against mixed states is given. The general amplitude
which describes the oscillation of two particles in the final states is
found. Using this EPR-type amplitude the problem of oscillation of particles
recoiling against kaons or neutrinos is resolved. The relativistic EPR
correlations on distances of the order of coherence lengths are considered.
\end{abstract}

\newpage
\section{Introduction}

The subject is not new. The problem is known since 1955 when Gell-Mann and
Pais [1] predicted the existence of two neutral kaons. Earlier, in 1953 in a
scheme for classifying the various newly-found particles, Gell-Mann
represented the neutral kaon $K^{0\mbox{ }}$and its antiparticle $\overline{K%
}^0$as two distinct particles. The decay of both particles into $\pi ^{+}\pi
^{-}$ was observed. If so, how do we know which particle has originated it:
the $K^{0\mbox{ }}$or the $\overline{K}^0?$ The problem has been solved by
realizing that what we observe is the mixture of two states, $K^{0\mbox{ }}$%
and $\overline{K}^0:$ 
\begin{equation}
\left| K_S\right\rangle =\frac 1{\left[ 2\left( 1+\left| \varepsilon \right|
^2\right) \right] ^{1/2}}\left[ \left( 1+\varepsilon \right) \left|
K^0\right\rangle +\left( 1-\varepsilon \right) \left| \overline{K}%
^0\right\rangle \right] , 
\end{equation}
\begin{equation}
\left| K_L\right\rangle =\frac 1{\left[ 2\left( 1+\left| \varepsilon \right|
^2\right) \right] ^{1/2}}\left[ \left( 1+\varepsilon \right) \left|
K^0\right\rangle -\left( 1-\varepsilon \right) \left| \overline{K}%
^0\right\rangle \right] , 
\end{equation}
where $\varepsilon $ is a small, complex, later measured parameter
responsible for CP symmetry breaking [2]. In this way first time the
interference between states of slightly different masses has appeared in
quantum mechanics. Inspired by the work of Gell-Mann and Pais, Bruno
Pontecorvo turned to consider the possibility of quantum mechanical mixing
in another neutral particle - the neutrino. In 1957 he first suggested that
a neutrino may oscillate into its antipartner [3]. Oscillation among the
different kinds of neutrinos was then proposed by Maki, Nakagawa and Sakata
in 1962 [4] and later by many others [5].

The neutral $K^{0\mbox{ }}-$ $\overline{K}^{0\mbox{ }}$boson system is not
the only one where the quantum mechanical mass mixing can be considered. We
can expect to observe the same phenomena in other neutral boson systems: $%
D^{0\mbox{ }}-$ $\overline{D}^{0\mbox{ }}$ and $B^{0\mbox{ }}-$ $\overline{B}%
^0.$ Generally, flavour oscillations of particles can occur when states
produced and detected in a given experiment, are superpositions of two or
more eigenstates with different masses. The oscillation of K and B meson
has been observed experimentally in 1961 [6] and later [7] and has been used
to place stringent constraints on physics beyond the Standard Model. If
neutrinos are massive and oscillate it is possible to resolve the well-known
solar neutrino problem [8]. There are also first experiments in which the
neutrino oscillations are observed [9].

The flavour oscillation of particles is a very fascinating demonstration of
quantum mechanics in the macroscopic world. It has served as a model for
many interesting systems and problems. Various aspects of quantum mechanics,
as for example coherence, decoherence, wave packets, measurements,
similarity and differences between pure and mixed states, wave function
collapse, EPR ''paradox'' are in action. On the other hand particle mixing
is the place where fundamental symmetries and properties of fundamental
interactions are studied. Discovering of the CP symmetry violation and the
measurement of differences between neutral mesons masses are connected with
K, B bosons mixing. Neutrino oscillations have a chance to be the first
place where problem of neutrino masses can be resolved.

In this review we will concentrate only on the quantum mechanical
description of particle oscillations. Problems connected with testing of the
fundamental interactions will not be discussed.

First of all we should mention that interference between states with
different masses is not allowed in non-relativistic quantum mechanics. The
Galilean invariance forbids a coherent superposition of such states (the so
called Bargman superselection rule [10]). Beyond the non-relativistic limit
such restrictions do not hold (which clearly follows from experiment). It
means that all of our considerations should be done in relativistic quantum
mechanics (nevertheless non-relativistic approximations are possible).

First, we would like to describe briefly the traditional approach to the
particle oscillation problem. This treatment is simple and elegant but
immediately raises a number of conceptual questions. We specify more of them
(Chapter 2). Next we show the wave packet treatment, where some of the
problems disappear (Chapter 3). The current density approach which is
closely connected with the experimental setting, is described in Chapter 4.
The problem of constructing the probability current density for a particle
with undetermined mass is also considered there.

Next, in Chapter 5, we give some remarks on the field theoretical approach
to particle oscillations. Usually (as in the case of neutrino oscillations)
the oscillating particle is not directly observed. Only particles
accompanying neutrinos, hadrons and charged leptons created in the decay are
observed. The proper approach should take all these circumstances into
account. The creation of the neutrino in the source, its propagation to the
detector and the detection process are treated in the framework of quantum
field theory as one large Feynman diagram.

In Chapter 6 we discuss the controversial problem of the oscillation of
particles recoiling against kaons or neutrinos from the production process.
A detailed approach using wave packets explains the problem of four-momentum
nonconservation raised in the literature.

In Chapter 7 we discuss the modern example of an Einstein-Podolsky-Rosen
correlation in K$^0-\overline{K}^0$and $B^{0\mbox{ }}-\overline{B}^0$
systems. The amplitude approach does not entail the somewhat mysterious
''collapse of the wave function'' which is usually invoked to describe the
EPR effects.

Finally in Chapter 8 we summarize our main conclusions.

\section{Problems connected with the traditional approach to the particle
oscillation.}

The usual description of kaon mixing phenomena can be found in many
textbooks [11]. Suppose, that we produce $K^{0\mbox{ }}$at $t=0$ by the
reaction 
\begin{equation}
\pi ^{-}p\rightarrow K^{0\mbox{ }}\Lambda ^0. 
\end{equation}
From (1) and (2) the $K^{0\mbox{ }}$state at $t=0$ is 
\begin{equation}
\left| K^0\right\rangle =\sqrt{\frac{1+\left| \varepsilon \right| ^2}{%
2\left( 1+\varepsilon \right) ^2}}\left( \left| K_S\right\rangle +\left|
K_L\right\rangle \right) . 
\end{equation}
After time t, as $\left| K_S\right\rangle $and $\left| K_L\right\rangle $
states are definite mass eigenstates, we have 
\begin{equation}
\left| K^0\left( t\right) \right\rangle =\sqrt{\frac{1+\left| \varepsilon
\right| ^2}{2\left( 1+\varepsilon \right) ^2}}\left( e^{-i\left( m_S-i\frac{%
\Gamma _S}2\right) t}\left| K_S\right\rangle +e^{-i\left( m_L-i\frac{\Gamma
_L}2\right) t}\left| K_L\right\rangle \right) 
\end{equation}
where $m_{L(S)}$ and $\Gamma _{L(S)}$ are masses and inverse mean lifetimes
respectively of the long (short)-lived component of K.

The $K^{0\mbox{ }}(\overline{K}^{0\mbox{ }})$ fraction of the beam after
time t is just 
\begin{eqnarray}
P_{K^0\rightarrow K^{0\mbox{ }}(\overline{K}^{0\mbox{ }})}\left( t\right)
&=&\left| \left\langle K^{0\mbox{ }}(\overline{K}^{0\mbox{ }})|K^{0\mbox{ }%
}(t)\right\rangle \right| ^2 \nonumber \\
&=&\frac 14\left[ e^{-\Gamma _St}+e^{-\Gamma
_Lt}\pm 2e^{-\frac 12(\Gamma _L+\Gamma _S)t}\cos \left( \Delta mt\right)
\right] 
\end{eqnarray}
where $\Delta m=m_L-m_S.$ From Eq.(6) we can see that the fraction of $K^{0%
\mbox{ }}(\overline{K}^{0\mbox{ }})$ becomes smaller (because of decay) and
changes with time with frequency $\omega =\left( \frac{\Delta m}{2\pi }\right) .$

Neutrino oscillations are described in a very similar way [12]. Let us
assume that at $t=0$ neutrino with flavour $\alpha $ was born with momentum $%
p$ perfectly defined (as for example neutrino $\nu _\mu $ in the pion decay $%
\pi ^{+}\rightarrow \mu ^{+}\nu _\mu $). At this time the neutrino state is
described by 
\begin{equation}
\left| \Psi _\alpha (0)\right\rangle =\sum_aU_{\alpha a}\left|
a\right\rangle , 
\end{equation}
where states $\left| a\right\rangle $ are energy-momentum eigenstates for
neutrinos with mass $m_a$ and $U_{\alpha a}$ are elements of a flavour-mass
mixing matrix. 
\\
Then 
\begin{equation}
H\left| a\right\rangle =E_a\left| a\right\rangle , 
\end{equation}
where $E_a=\sqrt{p^2+m_a^2}$ with the same momentum $p$ for each neutrino.
After time t the state will evolve into 
\begin{equation}
\left| \Psi _\alpha (0)\right\rangle \rightarrow \left| \Psi _\alpha
(t)\right\rangle =e^{-iHt}\left| \Psi _\alpha (0)\right\rangle
=\sum_aU_{\alpha a}e^{-iE_at}\left| a\right\rangle . 
\end{equation}
Then the probability that a neutrino born at $t=0$ with flavour $\alpha $ at
time t has flavour $\beta $ is given by 
\begin{equation}
P_{\alpha \rightarrow \beta }(t)=\left| \left\langle \Psi _\beta (0)|\Psi
_\alpha (t)\right\rangle \right| ^2=\left| \sum_{a=1}^nU_{\beta
a}^{*}e^{-iE_at}U_{\alpha a}\right| ^2, 
\end{equation}
where $n$ is the number of interfering light neutrinos\footnote
{To prove Eq.(10) we have to assume that the scalar product of two
eigenmomentum states $\left\langle b\mid a\right\rangle =\delta _{ab}.$
This
means that we must introduce some normalization volume and momentum and
energy are quantized.}.

Now usually relativistic approximations are made. As for real, light
neutrinos $p\gg m_a$, we have

(i) $E_a\cong p+\frac{m_a^2}{2p},$

and

(ii) a neutrino born in $x=0$, at time t will be approximately at position $%
x\approx t.$

Then from (10) we can find that the probability for our neutrino, born with
flavour $\alpha $, to have new flavour $\beta$ after traveling a distance $%
x,$ is 
\begin{equation}
P_{\alpha \rightarrow \beta }(x)=\sum_{a=1}^n|U_{\beta a}|^2|U_{\alpha
a}|^2+2\sum_{a>b}\left| U_{\beta a}^{*}U_{\alpha a}U_{\beta b}U_{\alpha
b}^{*}\right| \cos \left( 2\pi \frac x{L_{ab}}-\varphi _{ab;\alpha \beta
}\right) , 
\end{equation}
where $L_{ab},$ known as {\bf oscillation length} between $\nu _a$ and $\nu
_{b\mbox{ }}$is defined by 
\begin{equation}
L_{ab}=\frac{4\pi p}{m_a^2-m_b^2}, 
\end{equation}
and 
\begin{equation}
\varphi _{ab;\alpha \beta }=\arg \left( U_{\beta a}^{*}U_{\alpha a}U_{\beta
b}U_{\alpha b}^{*}\right) , 
\end{equation}
are phases responsible for CP violation.

From (11) it follows that the oscillation will disappear (the $P_{\alpha
\rightarrow \beta }(x)$ does not depend on $x$) if (i) all neutrino masses
are equal $m_a=m_b$ and/or (ii) only diagonal elements of the mixing matrix $%
U_{\alpha b}$ do not vanish.

The presented arguments seem to be clear and elegant but they are wrong.
Many conceptual questions arise when we look at the presentation shown
above. A complete treatment of particle oscillation must address the
following additional issues.
\\
(1) A necessary condition for particle oscillation to occur is that particle
source and detector are localized within the region $\Delta x$ much smaller
than the oscillation length $|L_{ab}|.$%
\begin{equation}
|L_{ab}|\gg \Delta x. 
\end{equation}
\\
(2) From Eqs. (4) and (7) we see that different mass eigenstates are
produced and detected coherently. This is possible only if the momentum $(p)$
and energy $\left( E\right)$ of the oscillating particle are spread in such
a way that the error in $m^2$ measurements given by 
\begin{equation}
\Delta m^2=\left[ \left( 2E\right) ^2\left( \Delta E\right) ^2+\left(
2p\right) ^2\left( \Delta p\right) ^2\right] ^{1/2} 
\end{equation}
is larger than $\left| m_a^2-m_b^2\right| \equiv |\Delta m_{ab}^2|,$%
\begin{equation}
\Delta m^2\geq  |\Delta m_{ab}^2| . 
\end{equation}
If this condition is not satisfied and $\left| m_a^2-m_b^2\right| \geq
\Delta m^2,$ then also 
\begin{equation}
\left| \Delta m_{ab}^2\right| \geq 2p\Delta p. 
\end{equation}
But from the uncertainty relation $\Delta x\geq \frac 1{\Delta p},$ and
Eq.(17) gives $\Delta x\geq \frac{2p}{\left| \Delta m_{ab}^2\right| }=\frac{%
|L_{ab}|}{2\pi },$ which is in contradiction with Eq.(14).

From both conditions described above we see that the oscillating particle
state cannot be described by a plane wave with definite momentum [13] and
the wave packet approach must be constructed.

(3) The energy and momentum conservation in processes in which oscillating
particles are created (e.g. $\pi ^{-}p\rightarrow \Lambda K^{0\mbox{ }}$or $%
\pi ^{+}\rightarrow \mu ^{+}\nu _\mu )$ implies that different mass
eigenstate components have different energy and momentum [14]. Approaches
where all oscillating particles have the same momentum and different
energies [11,12], or the same energies and different momenta [15, 16, 17]
are conceptually not correct.

(4) In the traditional approach to find the oscillation probability we
calculate the wave functions$^{^{\prime }}$ overlap (compare Eqs. (6) and
(10)). This procedure gives the probability which depends on time. In the
real experiment the distance between the source and the detector is known
(not the moment in which the measurement is done). To transform $P(t)$ into $%
P\left( x\right) $ the classical formula $x=vt$ is invoked. However to find
the probability that the beam of particles produced at $\stackrel{%
\rightarrow }{x}=0$ will reach a physical detector at a distance $\left| 
\stackrel{\rightarrow }{x}\right| $ the current density $\stackrel{%
\rightarrow }{j}(\stackrel{\rightarrow }{x},t)$ should be integrated over
the surface of the detector and over the time of observation [18] 
\begin{equation}
P\left (\left| \stackrel{\rightarrow }{x}\right|
,t_1<t<t_2 \right) 
=\int_{t_1}^{t_2}dt\int_{\partial A}\stackrel{\rightarrow }{dS}%
\cdot \stackrel{\rightarrow}{j}(\stackrel{\rightarrow}{x},t). 
\end{equation}
In such an approach there is no problem of how to change $t$ into $x$.

(5) In case of neutrinos additional conceptual problems arise because
neutrinos are not ''seen'' directly. The only things which can be ''seen''
are hadrons or/and charged leptons in points where neutrinos are produced
and detected. So in a realistic description, the external (initial and
final) particles should be described by wave packets, and the
mass-eigenstate neutrinos should propagate from the production region to a
detector [19, 20].

All points which have been mentioned above are not only purely academic. We
do not try to derive in a more precise way something which is known from the
beginning. We will see that the more precise approach to the particles$^{`}$
oscillation phenomenon gives us new predictions and elucidates mysteries in
many points. On the other hand we will see quantum mechanics in action on
macroscopic distances.

\section{The wave packet treatment of particle oscillation.}

The wave packet approach to neutrino oscillation was first proposed by B.
Kayser [13] and later considered in more detail in [21, 22]. Nowadays the
neutral boson oscillation is also treated in the same way. We will present
the formalism for neutrinos, but everything can be repeated also for bosons.

In Eq.(7) states $\left| a\right\rangle $ have definite energy and momentum
(also the spin direction of neutrinos is defined) so for the sake of
precision we should write 
\begin{equation}
\left| a\right\rangle \equiv \left| a;p\right\rangle =\left| a;\stackrel{%
\rightarrow }{p},E,\sigma \right\rangle 
\end{equation}
We can easily construct a state with momentum distributed around a mean
value $\stackrel{\rightarrow }{p_a}$. Let us assume that, instead of a
plane wave, our new state $\left| a\right\rangle $ (Eq.(20)) has a Gaussian
form, which in the momentum representation, is given by 
\begin{equation}
\left\langle a;\stackrel{\rightarrow }{p}\mid a\right\rangle \equiv \Psi
_a\left( \stackrel{\rightarrow }{p},\stackrel{\rightarrow }{p}_a,\sigma
_{pP}\right) =\frac 1{\left[ \sqrt{2\pi }\sigma _{pP}\right] ^{3/2}}\exp
\left[ -\frac{\left( \stackrel{\rightarrow }{p}-\stackrel{\rightarrow }{p}_a%
\right) ^2}{4\sigma _{pP}^2}\right] , 
\end{equation}
where the width $\sigma _{pP}$ is the same for each massive neutrino in the
production $(P)$ process and the same along all three directions. The
average momenta $\stackrel{\rightarrow }{p_a}$ of the different mass
eigenstates are determined by the kinematics of the production process.

In the wave packet approach, the flavour states $\left| \Psi _\alpha
(t)\right\rangle $ after time t (given by Eq.(9) in the plane wave
formalism) are now 
\begin{eqnarray}
\left| \Psi _\alpha (t)\right\rangle &=&\sum\limits_aU_{\alpha a}e^{-iHt}\int
d^3p\left| a; 
\stackrel{\rightarrow }{p}\right\rangle \left\langle a;\stackrel{\rightarrow 
}{p}\mid a\right\rangle \nonumber  \\ 
&=&\sum\limits_aU_{\alpha a}\int d^3p \Psi _a\left( \stackrel{\rightarrow }{p},%
\stackrel{\rightarrow }{p}_a,\sigma _{pP}\right) e^{-iE_a(p)}\left| a;%
\stackrel{\rightarrow }{p}\right\rangle , 
\end{eqnarray}
where $E_a \left( \vec{p} \right)  =\sqrt{p^2+m_a^2}.$

The same states in the position representation $\left| b;\stackrel{%
\rightarrow }{x}\right\rangle $ are given by 
\begin{equation}
\left| \Psi _\alpha (t)\right\rangle =\sum\limits_b\int d^3x\left| b; 
\stackrel{\rightarrow }{x}\right\rangle \left\langle b;\stackrel{\rightarrow 
}{x}\mid \Psi _a(t)\right\rangle =
\sum\limits_aU_{\alpha a}\int d^3x\Psi _a\left( \stackrel{\rightarrow }{x},t;%
\stackrel{\rightarrow }{v_a},\sigma _{xP}\right) \left| a;\stackrel{%
\rightarrow }{x}\right\rangle , 
\end{equation}
where now the function $\Psi _a\left( \stackrel{\rightarrow }{x},t;\stackrel{%
\rightarrow }{v_a},\sigma _{xP}\right)$ is defined by 
\begin{eqnarray}
\Psi _a\left( 
\stackrel{\rightarrow }{x},t;\stackrel{\rightarrow }{v_a},\sigma
_{xP}\right)& =& \int d^3p\Psi _a\left( \stackrel{\rightarrow }{p},\stackrel{%
\rightarrow }{p}_a,\sigma _{pP}\right) e^{-iE_a(p)t}\left\langle a;\stackrel{%
\rightarrow }{x}\mid a;\stackrel{\rightarrow }{p}\right\rangle \nonumber  \\ 
&=&\frac 1{\left( 2\pi \right) ^{3/2}}\int d^3p\Psi _a\left( \stackrel{%
\rightarrow }{p},\stackrel{\rightarrow }{p}_a,\sigma _{pP}\right) e^{i\left( 
\stackrel{\rightarrow }{p}\stackrel{\rightarrow }{x}-E_a(p)t\right) }. 
\end{eqnarray}
Since the Gaussian wave packet in momentum space is picked around the
average momentum $\stackrel{\rightarrow }{p_a}$ we can neglect the spreading
of the wave packet and approximate 
$$
E_a(p)=E_a+\stackrel{\rightarrow }{v_a}\left( \stackrel{\rightarrow }{p}-%
\stackrel{\rightarrow }{p}_a\right) 
$$
where 
\begin{equation}
E_a=\sqrt{p_a^2+m_a^2}\mbox{ and }\stackrel{\rightarrow }{v_a}=\frac{%
\partial E_a}{\partial \stackrel{\rightarrow }{p}}|_{\stackrel{\rightarrow }{%
p}=\stackrel{\rightarrow }{p_a}}=\frac{\stackrel{\rightarrow }{p_a}}{E_a}. 
\end{equation}
Then 
\begin{equation}
\Psi _a\left( \stackrel{\rightarrow }{x},t;\stackrel{\rightarrow }{v_a}%
,\sigma _{xP}\right) =\frac 1{\left[ \sqrt{2\pi }\sigma _{xP}\right]
^{3/2}}\exp \left[ i\left( \stackrel{\rightarrow }{p_a}\stackrel{\rightarrow 
}{x}-E_at\right) -\frac{\left( \stackrel{\rightarrow }{x}-\stackrel{%
\rightarrow }{v_a}t\right) ^2}{4\sigma _{xP}^2}\right] , 
\end{equation}
with the width $\sigma _{xP}$ in coordinate space given by 
\begin{equation}
\sigma _{xP}=\frac 1{2\sigma _{pP}}. 
\end{equation}
As earlier in Eq.(10), to find the amplitude of the flavour changing
process, we project the states $\left| \Psi _\alpha (t)\right\rangle $ on
the flavour states $\left| \Psi _\beta (0)\right\rangle $%
\begin{equation}
A_{\alpha \rightarrow \beta }(t)=\left\langle \Psi _\beta (0)\mid \Psi
_\alpha (t)\right\rangle . 
\end{equation}
If $\left| \Psi _\beta (0)\right\rangle $ is the same as before (Eq.(7))
this means that the momentum of each neutrino $v_a$ is measured precisely.
But this is not realistic, so let us assume that also detection process is
characterized by the spatial coherence width $\sigma _{pD}$ connected with
the uncertainties in momentum and energy measurements 
\begin{equation}
\left| \Psi _\beta (0)\right\rangle =\sum_bU_{\beta b}\int d^3p\Psi _b\left( 
\stackrel{\rightarrow }{p},\stackrel{\rightarrow }{p}_b,\sigma _{pP}\right)
\left| b;\stackrel{\rightarrow }{p}\right\rangle . 
\end{equation}
The average values of the momentum $\stackrel{\rightarrow }{p_b}$ are the
same as in the incoming wave packets Eq.(22). To calculate the spatial
decomposition of the detecting flavour state we have to take into account
that the detector is placed at a distance $L$ from
the origin of the coordinates, so we have 
\begin{equation}
\left| \Psi _\beta (\stackrel{\rightarrow }{L})\right\rangle =\sum_b\int
d^3x_D\left| b,\stackrel{\rightarrow }{x_D}\right\rangle \left\langle b,%
\stackrel{\rightarrow }{x_D}|\Psi _\beta (0)\right\rangle , 
\end{equation}
and after the same approximation as before (Eq.(24)) we obtain 
\begin{equation}
\left| \Psi _\beta (\stackrel{\rightarrow }{L})\right\rangle =\sum_bU_{\beta
b}\int d^3x_D\Psi _b\left( \stackrel{\rightarrow }{x_D},0;\stackrel{%
\rightarrow }{v_b},\sigma _{xD}\right) \left| b,\stackrel{\rightarrow }{x_D}%
\right\rangle .
\end{equation}
$\Psi _b\left( \stackrel{\rightarrow }{x_D},0;\stackrel{\rightarrow }{%
v_b},\sigma _{xD}\right) $ is given exactly by Eq.(25) after replacements 
\begin{equation}
a\rightarrow b,\;\;\stackrel{\rightarrow }{x}\rightarrow \stackrel{\rightarrow }{%
x_D},\;\;t\rightarrow 0,\;\;\sigma _{xP}\rightarrow \sigma _{xD}. 
\end{equation}
The amplitude of the flavour changing process is given by the overlap 
\begin{equation}
A_{\alpha \rightarrow \beta }(\stackrel{\rightarrow }{L},t)=\left\langle
\Psi _\beta (\stackrel{\rightarrow }{L})|\Psi _\alpha (t)\right\rangle .
\end{equation}
We have to remember that the origin of coordinates $\stackrel{%
\rightarrow }{x_D}$ and $\stackrel{\rightarrow }{x}$ are not the same, so 
\begin{equation}
\left\langle b,\stackrel{\rightarrow }{x_D}|a,\stackrel{\rightarrow }{x}%
\right\rangle =\delta _{ab}\delta ^{(3)}\left( \stackrel{\rightarrow }{x_D}+%
\stackrel{\rightarrow }{L}-\stackrel{\rightarrow }{x}\right) , 
\end{equation}
and we have 
\begin{eqnarray*}
A_{\alpha \rightarrow \beta }( 
\stackrel{\rightarrow }{L},t)&=&\sum\limits_aU_{\beta a}^{*}U_{\alpha a}\int d^3x\Psi
_a^{*}\left( \stackrel{\rightarrow }{x}-\stackrel{\rightarrow }{L},0;%
\stackrel{\rightarrow }{v_a},\sigma _{xD}\right) \Psi _a\left( 
\stackrel{\rightarrow }{x},t;\stackrel{\rightarrow }{v_a},\sigma
_{xP}\right)  \\ 
&=& \sqrt{\frac{2\sigma _{xP}\sigma _{xD}}{\sigma _x^2}}\sum\limits_aU_{\beta
a}^{*}U_{\alpha a}\exp \left[ -i\left( E_at-\stackrel{\rightarrow }{p_a}%
\stackrel{\rightarrow }{L}\right) -\frac{\left( \stackrel{\rightarrow }{L}-%
\stackrel{\rightarrow }{v_a}t\right) ^2}{4\sigma _x^2}\right] , 
\end{eqnarray*}
\begin{equation}
\end{equation}
where the total production and detection width is now 
\begin{equation}
\sigma _x=\sqrt{\sigma _{xP}^2+\sigma _{xD}^2}. 
\end{equation}
Everything we have done up to now can be applied either to neutrino or
neutral boson oscillations. The only assumption about the narrow wave
packets in the momentum representation (Eq.(24)) can be used in both cases.

Next we calculate the oscillation probability for neutrinos which are
relativistic $(p\gg m)$ and following Ref.[22] we approximate 
\begin{equation}
E_a\cong E+\xi \frac{m_a^2}{2E},\mbox{ }p_a\cong E-\left( 1-\xi \right) 
\frac{m_a^2}{2E}, 
\end{equation}
and 
$$
v_a\cong 1-\frac{m_a^2}{2E^2}. 
$$
E is the energy determined by kinematics of the production process for
a massless neutrino and $\xi $ is a dimensionless quantity of order unity.
We will see (next Chapter) how the relativistic approximation for neutrinos
causes that the production and detection processes can be factorized out and
the standard quantum mechanical approach describes the oscillation
phenomenon properly.

In all realistic experiments the distance $L$ is a fixed and known quantity, whereas time t is not measured. The
quantity which we measure is the time integral of the probability. Now the
time integral can be done, and it is possible to avoid the not properly
legitimated (in quantum mechanics) replacement $x=vt.$

The time integral for $\left| A_{\alpha \rightarrow \beta }(\stackrel{%
\rightarrow }{L},t)\right| ^2$ (Eq.(34)) can be done [22] and after
normalization $\left( \sum_\beta P_{\alpha \rightarrow \beta }(x)=1\right)$%
, instead of Eq.(11) we have 
\begin{eqnarray}
P_{\alpha \rightarrow \beta }(x)&=&\int_0^\infty \left| A( 
\stackrel{\rightarrow }{L},t)\right| ^2dt \nonumber \\
&=&\sum\limits_a\left| U_{\beta b}\right|
^2\left| U_{\alpha a}\right| ^2+ 
2\sum\limits_{a>b}\left| U_{\beta b}^{*}U_{\alpha a}U_{\beta b}U_{\alpha
b}^{*}\right| \cos \left( 2\pi \frac x{L_{ab}^{osc}}-\varphi _{ab;\alpha
\beta }\right) \nonumber \\
&\times& e^{-\left( \frac x{L_{ab}^{coh}}\right) ^2}
e^{- {\left(  
\frac{\scriptstyle x}{\scriptstyle L_{ab}^{coh}} \right)}^2} 
e^{-2\pi^2 \xi^2 {\left(  
\frac{\scriptstyle \sigma_x}{\scriptstyle L_{ab}^{osc}} \right)}^2}. 
\end{eqnarray}
The oscillation lengths are the same as before (Eq.(12)), namely 
\begin{equation}
L_{ab}^{osc}=\frac{4\pi E}{\Delta m_{ab}^2},\;\;\Delta m_{ab}^2=m_a^2-m_b^2, 
\end{equation}
and $L_{ab}^{coh}$ known as {\bf coherence lengths} [23] are given by 
\begin{equation}
L_{ab}^{coh}=\frac{4\sqrt{2}\sigma _xE^2}{|\Delta m_{ab}^2|}. 
\end{equation}
Comparing Eq.(37) to the usual expression for the neutrino oscillation
probability we can see that two additional terms appear.

The second factor 
\begin{equation}
e^{-2\pi^2 \xi^2 {\left(  
\frac{\scriptstyle \sigma_x}{\scriptstyle L_{ab}^{osc}} \right)}^2} 
\end{equation}
is equal to unity if $\sigma _x\ll \left| L_{ab}^{osc}\right| .$ This
inequality must be satisfied to observe any oscillation. The presence of the
term (40) which goes to zero for $\sigma _x>\left| L_{ab}^{osc}\right| ,$
reflects the requirement which we qualitatively discussed in the previous
Chapter: to see the oscillations, the localization of the source and the
detector must be much better determined than the oscillation length.

The first factor 
\begin{equation}
e^{- {\left(  
\frac{\scriptstyle x}{\scriptstyle L_{ab}^{coh}} \right)}^2} 
\end{equation}
was predicted long ago [23]. It is connected with the fact that two wave
packets each with different momentum and energy, have slightly different
group velocities. It means that after some time the mass eigenstate wave
packets no longer overlap and cannot interfere to produce oscillations. It
is very easy to predict the value of coherence length. If both wave packets
have width $\sigma _{xP}$ along the direction of propagation and the
difference between group velocities is $\left| v_a-v_b\right| =\Delta v$
then we can expect that after traveling a distance L 
\begin{equation}
L=\frac{2\sigma _{xP}}{\Delta v}\frac{v_a+v_b}2 
\end{equation}
both wave packets cease to overlap each other. This L is just the coherence
length.

For relativistic neutrinos, using approximations given by Eq.(36), we
reproduce $L_{ab}^{coh}$ from Eq.(39), to a factor of $\sqrt{2}.$ We can
expect that the coherence length becomes longer if the spreading of the wave
packets is taken into account. It is indeed the case as it was proved in
Ref.[21].

From Eq.(39) we see that the coherence length $L_{ab}^{coh}$ is
proportional to $\sigma _x=\sqrt{\sigma _{xP}^2+\sigma _{xD}^2}$ and not
only to $\sigma _{xP}$ as in Eq.(42). It means that precise measurements of
momenta of all particles appearing in the detection process (which implies
small $\sigma _{pD}$, thus large $\sigma _{xD}$) can increase the coherence
length [22, 24]. This is a wonderful example of quantum mechanics in action.
A measurement can restore the coherence. Two wave packets having negligible
overlap in the detector (thus without detector influence they cannot
interfere, and the oscillation disappears, $\sigma _x=\sigma _{xP}$),
because of precise measurements $\left( \sigma _{xD}\gg \sigma _{xP}\right)
, $ may still interfere to give rise to oscillations $\left( \sigma _x\gg
\sigma _{xP}\right) $. This feature of quantum mechanics disagrees with
causality. However, it is not the first time when quantum mechanics is at
variance with common sense.

But Eq.(37) also restores some common sense. Measurements of momenta and
energies of detected particles cannot be too precise if we want to maintain
the particle oscillation. As a matter of fact, we have a longer and longer
coherence length, but on the other hand increasing $\sigma _x$ makes the
position of the detector to be more and more undefined. If $\sigma _x>\left|
L_{ab}^{osc}\right| $ the wave packets of neutrinos $\nu _a$ and $\nu _b$
lose coherence (Eq.(40)), the oscillation between them is wash away. We see,
particularly, that the plane wave approach and oscillations are
incompatible. For a plane wave $\sigma _x\rightarrow \infty $ $\left( \sigma
_p=0\right) $ and the factor (40) teaches us that oscillations disappear.

There are also approaches where the neutrino oscillation is treated in a
manifestly Lorentz invariant way [25] . The final answer is exactly the same
as was presented up to now, but for one difference. In the fully covariant
treatment, besides the spatial width $\sigma _x,$ also temporal width $%
\sigma _t$ should appear. It causes only one change in the oscillation
probability formula Eq.(37). Instead of the spatial width $\sigma _x,$ a new
effective one, $\overline{\sigma }_{ab}$ appears in the coherence length
(39) and in the factor (40) 
\begin{equation}
\sigma _x\rightarrow \overline{\sigma }_{ab}=\sigma _x+\frac{v_a^2+v_b^2}{%
v_a+v_b}\sigma _t 
\end{equation}
where $v_a$ and $v_b$ are group velocities of wave packets.

If $\sigma _t$ is given by the lifetime of particles which produce neutrino, 
$\sigma _t$ = $\tau $ $[25]$ (e.g. pions, kaons or muons) then the second
term in (43) is usually much bigger than the first one and the role of
factor (40) becomes more important. We should stress however that the
independent widths of spatial and temporal characteristics of wave packets
cause that freely propagating particles are not necessary on mass-shell. If
we insist to have our particle on mass shell (all the time, not for mean
values exclusively, which is equivalent with the requirement that our
particle's state satisfies the equation of motion) only the
momentum (or only the energy) distribution should be applied. Then energy
(or momentum) is distributed also but in agreement with the on-shell
relation $E=\sqrt{p^2+m^2}.$ Such an approach, which we also used in our
presentation above is the standard one.

\section{Current density approach to particle oscillations and the problems
of states with undefined masses.}

A typical experiment which tries to observe particle oscillations measures
the flux of flavour $\beta $ type particles in the detector localized at
some distance L from the source which produces particles with flavour $%
\alpha .$ The time of measurements is not known. Usually typical
measurements last hours, days or even years (like the observation of solar
neutrinos). So the most appropriate way to find the probability (or number
of particles) to cross the surface $\partial A$ of the detector (see Fig.
(1)) is to integrate the probability current density over the surface
and integrate
the result once more over the duration of measurements 
\begin{equation}
P_{\alpha \rightarrow \beta }(L)=\int_{t_1}^{t_2}dt\int_{\partial A}%
\stackrel{\rightarrow }{dS} \cdot \stackrel{\rightarrow }{j_\beta }(\stackrel{%
\rightarrow }{x},t). 
\end{equation}

This procedure seems to be so easy and natural, that we can ask why people
try to use other, more complicated methods. The answer is very simple, there
is a problem with the correct definition of the current $j_\beta (\stackrel{%
\rightarrow }{x},t).$ This current should be defined for particles which we
measure, that is $K^0,\overline{K}^0,\nu _e$ or $\nu _\mu $. But these
particles have undefined masses, and we do not know how to define the
probability current for such particles. The problem is more general: how to
define properly the creation and annihilation operators for undefined mass
states [26,27]? Here we will not discuss all trials [26] to resolve this
problem in quantum field theory. We will concentrate only on a simple
example of definition of the current $j_\beta $ for particles with flavour $%
\beta .$ For relativistic currents the problem has not been solved. For
kaons it was done in Ref.[18]. In the non-relativistic case the free
Schr\"{o}dinger equation for particles with mass $m_a$ can be written in the
form 
\begin{equation}
i\frac{\partial \Psi _a}{\partial t}=\left( -\frac \Delta {2m_a}+m_a \right)
\Psi _a, 
\end{equation}
which is the appropriate non-relativistic limit of either Klein-Gordon or
Dirac equations $(E\approx \frac{p^2}{2m}+m)$. For the Schr\"{o}dinger equation
(Eq.(45)) we know how to define the probability current: 
\begin{equation}
\stackrel{\rightarrow }{j_a}(\stackrel{\rightarrow }{x},t)=\frac
1{m_a}Im(\Psi _a^{*}(\stackrel{\rightarrow }{x},t)\stackrel{\rightarrow }{%
\nabla }\Psi _a(\stackrel{\rightarrow }{x},t)) 
\end{equation}
for which the usual continuity equation is satisfied 
\begin{equation}
\frac \partial {\partial t}(\Psi _a^{*}\Psi _a)+div\stackrel{\rightarrow }{%
j_a}=0. 
\end{equation}
The problem arises while we try to define the current for the states 
\begin{equation}
\Psi _\alpha =c\Psi _a+d\Psi _b\mbox{ and }\Psi _\beta =-d^{*}\Psi _a+c\Psi
_b\mbox{ , }\left| c\right| ^2+\left| d\right| ^2=1 
\end{equation}
which are an orthogonal mixture of two states with different masses $m_a$
and $m_b$. We expect that, because of mass mixing, the currents for $\Psi
_{\alpha ,\beta }$ will not be conserved [18], so let us propose a modified
''continuity equation'' for $\Psi _{\alpha ,\beta }$ states in the form 
\begin{equation}
\frac \partial {\partial t}|\Psi _{\alpha ,\beta }|^2+div\stackrel{%
\rightarrow }{j_{\alpha ,\beta }}=d_{\alpha ,\beta }. 
\end{equation}
With the following requirements, concerning the new current $\stackrel{%
\rightarrow }{j_{\alpha ,\beta }}$[18]$,$ that: (1) only ''velocity'' terms
with one gradient are included, (2) for $m_a$ $\rightarrow $ $m_b.$ the
''diffusion'' terms $d_{\alpha ,\beta }$ should vanish $d_{\alpha ,\beta
}\rightarrow 0$ , and (3) the sum of both flavour currents $\stackrel{%
\rightarrow }{j_\alpha }+\stackrel{\rightarrow }{j_\beta }$ is conserved 
\begin{equation}
\frac \partial {\partial t}(|\Psi _\alpha |^2+|\Psi _\beta |^2)+div(%
\stackrel{\rightarrow }{j_\alpha }+\stackrel{\rightarrow }{j_\beta })=0, 
\end{equation}
the currents $\stackrel{\rightarrow }{j_{\alpha ,\beta }}$ and the diffusion
terms $d_{\alpha ,\beta }$ can be found. 
\begin{equation}
\stackrel{\rightarrow }{j_\alpha }=\left| c\right| ^2\stackrel{\rightarrow }{%
j_a}+\left| d\right| ^2\stackrel{\rightarrow }{j_b}+Im\left[ cd^{*}(\frac
1{m_a}\Psi _b^{*}grad\Psi _a-\frac 1{m_b}\Psi _a^{*}grad\Psi _b^{*})\right]
, 
\end{equation}
and 
\begin{equation}
d_\alpha =\left( m_a-m_b\right) Im\left[ cd^{*}(2\Psi _a\Psi _b-\frac
1{m_am_b}(grad\Psi _a)(grad\Psi _b^{*}))\right] . 
\end{equation}
For $\stackrel{\rightarrow }{j_\beta }$ and $d_\beta $ we have 
\begin{equation}
\stackrel{\rightarrow }{j_\beta }=\stackrel{\rightarrow }{j_\alpha }\left(
c\rightarrow -d^{*},d\rightarrow c^{*}\right) ,\mbox{and }d_\beta =d_\alpha
\left( c\rightarrow -d^{*},d\rightarrow c^{*}\right) . 
\end{equation}
Calculations for $K^0-\overline{K}^0$mixing probability, using the
definition (Eq.(44)) have been done in Ref.[18]. The standard formula
(Eq.(6)), with the same oscillation frequency $\omega =\frac{\Delta m}{2\pi }
$ was recovered, supporting previous results.

There is, however, one objection concerning this approach. The flavour
currents which we use are not conserved. This nonconservation is given by
the diffusion term (Eq.(52)) which is proportional to $\Delta
m_{ab}=m_a-m_b $. So, we can expect that all our calculations have been done
with the some precision. Then, if the result is proportional to $\Delta
m_{ab},$ our probability flux calculations give the correct answer. The
diffusion terms will change the result in higher powers of $\Delta m_{ab}.$

\section{Treatment of particle oscillation in the framework of quantum field
theory.}

There are several papers where authors investigate the neutrino oscillation
problem in the framework of quantum field theory [19,20,24,28,29,30]. What
are the main reasons for those studies?

First until now, to describe particle oscillations, we have used states with
undefined masses (Eq.(4) for kaons and Eq.(7) for neutrinos). But, as we
have seen in the previous Chapter, there is a problem of proper definition
of such states [26]. Only in the extremely relativistic limit the flavour
states are defined correctly [27]. As we will see, in quantum field theory
the particle oscillation can be treated without resort to weak eigenstates.

Secondly, we have completely neglected the effect of the production and
detection processes. It has been shown [27], that the neutrino oscillation
probability is independent from the details of the production and detection
processes only in the case of extremely relativistic neutrinos.

And finally, in real neutrino oscillation experiments only associated
particles, hadrons and charged leptons are observed. Neutrinos are not
prepared and not observed directly. One can measure the energy and momentum
distributions of other particles which appear in the production and
detection processes. As we will see, only the quantum field theoretical
approach gives the opportunity to express the neutrino oscillation
probability in terms of measured quantities.

Let us now describe briefly how the particle oscillation is treated in field
theory. As in Ref.[20,30] we will describe the process of neutrino
production (P) and neutrino detection (D) as one Feynman diagram with a
virtual neutrino propagating itself on macroscopic distances between the
source and the detector. Let us consider the process [20,30] 
\begin{equation}
P_I\rightarrow P_F+l_\alpha ^{+}+\nu _\alpha \stackrel{\left( \nu _\alpha
\rightarrow \nu _\beta \right) }{\rightarrow }\nu _\beta +D_I\rightarrow
D_F+l_\beta ^{-} 
\end{equation}
where $P_I$ and $P_F$ ($D_I$ and $D_F$) are the particles in the production
(detection) processes.

The production and detection processes are localized in coordinates $%
\stackrel{\rightarrow }{x_P}(\stackrel{\rightarrow }{x_D})$ and times $%
t_P(t_D)$ (see Fig.2). All initial $\left( P_I,P_F,l_\alpha ^{+}\right) $
and final $\left( D_I,D_F,l_\beta ^{-}\right) $ particles are described by
wave packets. Their shapes depend on the measurement precision in the
production and detection processes. The amplitudes for the full process can
be written in the form 
\begin{equation}
A_{P\rightarrow D}=\left\langle P_F,l_\alpha ^{+};D_F,l_\beta
^{-}|S|P_ID_I\right\rangle . 
\end{equation}
We can see that there are no neutrinos in the initial and final states. Only
particles which really appear in the production and detection ''equipments''
are observed. Neutrinos with mass $m_{a\mbox{ }}$propagate virtually between
the source and the detector and are described by the Feynman propagators 
\begin{equation}
\left\langle 0|T\left( \nu _a\left( x_1\right) \nu _b\left( x_2\right)
\right) |0\right\rangle =\delta _{ab}\int \frac{d^4k}{\left( 2\pi \right) ^3}%
\frac{\widehat{k}+m_a}{k^2-m_a^2+i\varepsilon }e^{-ik\left( x_1-x_2\right)
}. 
\end{equation}
We will not present the details of all calculations, which are
straightforward but tedious. A clear presentation can be found in Refs. [20,
28, 30]. We will concentrate only on the discussion of the final results.

First of all, neutrinos are not directly present in Eq.(55), but this is not
necessary. The amplitude $A_{P\rightarrow D}$ depends on points $\left( 
\stackrel{\rightarrow }{x_P},t_P\right) $ and $\left( \stackrel{\rightarrow 
}{x_D},t_D\right) $ where neutrinos were born and detected, and this is
enough to study oscillations.

Next, the amplitude $A_{P\rightarrow D}$ depends also on amplitudes of the
production and detection processes and the full structure of $%
A_{P\rightarrow D}$ is the following 
\begin{equation}
A_{P\rightarrow D}=\sum_aU_{\beta a}^{*}U_{\alpha a}\mbox{ }A_af_a\left( 
\stackrel{\rightarrow }{x_D}-\stackrel{\rightarrow }{x_P},t_D-t_P\right) , 
\end{equation}
where $A_a$ describes the process of neutrino creation and annihilation. The
standard oscillation formula is recovered only if $A_a$ can be factorized.
This happens, when amplitudes $A_a$ become independent of neutrino masses, $%
A_a=A$ . If all neutrinos are relativistic then $A_a=A\left( m_a\cong
0\right) $ and the oscillation probability can be defined.

In case of relativistic neutrinos the time integrated neutrino flavour
changing probability is given by the similar formula to Eq.(37) with two
changes. First, the dumping factor (Eq.(40)) is slightly modified and now
equals 
\begin{equation}
e^{-2\pi^2\omega \xi {\left(  
\frac{\scriptstyle \sigma_x}{\scriptstyle L_{ab}^{osc}} \right)}^2} 
\end{equation}
where $\xi $ is the same quantity as before (Eq.(36)), but $\omega $ is the
new factor which depends on the production and detection dynamics 
and can be large (e.g. $\omega \cong 10$ is possible [30]). The
second modification is a little different definition of the coherence
length. Instead of (Eq.(39)) we now have 
\begin{equation}
L_{ab}^{osc}=\sqrt{2\omega }\frac{4E^2\sigma _x}{\left| \Delta
m_{ab}^2\right| }, 
\end{equation}
with the same factor $\omega $ as in Eq.(58).

The oscillation length $L_{ab}^{osc}$ and the spatial width $\sigma _x$ are
given by the same formulae as before (Eqs.(38) and (35)).

There is also an additional very important difference between the present
and the former wave packet approaches. Before, $\sigma _{xP}$ and $\sigma
_{xD}$ (Eq.(35) were two spatial widths of neutrinos specified in some way
by the production and detection processes respectively. Now, these
quantities are defined by spatial widths of hadrons and leptons which are
measured. It turns out (for detail see Ref.[30])

$$
\frac 1{\sigma _{xP}^2}=\frac 1{\sigma _{xP_I}^2}+\frac 1{\sigma
_{xP_F}^2}+\frac 1{\sigma _{x\alpha }^2}, 
$$
and 
\begin{equation}
\frac 1{\sigma _{xD}^2}=\frac 1{\sigma _{xD_I}^2}+\frac 1{\sigma
_{xP_F}^2}+\frac 1{\sigma _{x\beta }^2}. 
\end{equation}
The widths of observed particles in the production and detection processes
define the width of '' the neutrino'', even if there is no place in this
approach for physical neutrinos (only virtual ones appear). In the
configuration space (Eq.(60)) the sum of the inverse squares of widths for
the observed particles gives the inverse square of the resultant width. Then
the smallest ingredient width dominates the values of $\sigma _{xP}$ or $%
\sigma _{xD}$. It is the opposite for the resultant width $\sigma _x,$ where 
$\sigma _x^2$ $=\sigma _{xP}^2+\sigma _{xD}^2.$ From the definition of the
momentum width $\sigma _p=\frac 1{2\sigma _x}$ , it follows that it is just
opposite in momentum space, so then 
\begin{equation}
\frac 1{\sigma _p^2}=\frac 1{\sigma _{pP}^2}+\frac 1{\sigma _{pD}^2}, 
\end{equation}
but 
\begin{equation}
\sigma _{pP}^2=\sigma _{pP_I}^2+\sigma _{pP_F}^2+\sigma _{p\alpha }^2, 
\end{equation}
and the same for particles in the detection process. We can see explicitly,
that precise measurements of momenta of all particles involved in the
neutrino detection process (small $\sigma _{pD_I},\sigma _{pD_F}$ and $%
\sigma _{p\alpha }$) give a small resultant width $\sigma _{pD}$, thus large 
$\sigma _{xD}$ and large $\sigma _x$. The same subtle thing, which we have
discussed before in Chapter 3, that the final measurement is able to recover
the interference, in the present interpretation has found a much stronger
background.

\section{Do particles recoiling against mixed states oscillate?}

For many years oscillations of particles like kaons or neutrinos were
treated in isolation. The circumstances in which oscillating particles were
produced have not been considered. Recently a series of papers has appeared
[31, 32], in which the kinematics of the production process have been taken
into account in detail. Authors claim that, because the produced oscillating
particles have neither momentum nor energy defined in explicit way, this
fact should have consequences not only for them but also for the recoiling
particles.

Let us consider the $K^0$ production in the reaction 
\begin{equation}
\pi ^{-}p\rightarrow \Lambda K^0, 
\end{equation}
or the neutrino production in the $\pi ^{+}$ decay 
\begin{equation}
\pi ^{+}\rightarrow \mu ^{+}\nu _\mu . 
\end{equation}
If the invariant mass of the initial system is denoted by M $(M^2=(P_{\pi
^{-}}+P_p)^2$ or $M^2=m_\pi ^2)$ then energies $(E_i)$ and momenta $\left(
p_i\right) $ of outgoing particles depend on masses $\left( m_i\right) $ of $%
K^0$ or $\nu _\mu $. In the CM system there is 
$$
p_i=\frac{\left[ \left( M^2-m_i^2-m^2 \right)^2-4m_i^2m^2 \right]
^{1/2}}{2M}, 
$$
and 
\begin{equation}
E_i=\frac{M^2+m_i^2-m^2}{2M}, 
\end{equation}
where $m$ is the mass of the recoiling particle $\Lambda $ or $\mu ^{+}$.
From the four-momentum conservation in the production processes (63) or
(64), the energy and momentum of $\Lambda $ or $\mu ^{+}$ are also defined
by (65). 
Authors of Ref. [31] claim that if $K^0$ or $\nu
_\mu $ oscillate, also the recoiling particles $\Lambda $ or $\mu ^{+}$ do
the same. If it is true, the existence of such phenomena could give a chance
for indirect observation of neutrino oscillations which are very difficult
to observe in a direct way. However other papers have immediately appeared
[33, 34, 35] where authors have been strongly against the oscillation of
particles produced in association with kaons or neutrinos. We will present
here our approach to the problem [36] which also supports the opinion
against a visible oscillation of the associated particles. To fix notations,
everything will be described for the process (63), but equally well we can
show the lack of visible muon oscillations in the pion decay (64).

First of all we would like to specify the kind of oscillation, we can
consider for $\Lambda $ (or $\mu ^{+}$). Two $\Lambda $'s with different
masses do not exist. But even without mass differences the $\Lambda $'s are
produced in association with the long-live $K_L$ and the short-live $K_S.$
As $K_L$ and $K_S$ have different masses $\Lambda $'s will be produced in
two orthogonal states with different energy and momentum. 
$$
\left| \Lambda _L\right\rangle =\left| -\stackrel{\rightarrow }{p_L}%
,M-E_L\right\rangle 
$$
and 
\begin{equation}
\left| \Lambda _S\right\rangle =\left| -\stackrel{\rightarrow }{p_S}%
,M-E_S\right\rangle 
\end{equation}
where $\overrightarrow{p}_{L(S)}$ and $E_{L(S)}$ are the momentum and the
energy of the $K_L(K_S)$ in total CM frame of the process (63). We do not
know in which state $\left| \Lambda _L\right\rangle $ or $\left| \Lambda
_S\right\rangle $ the $\Lambda $ particles are produced, so let us assume
that at $t=0$ they are produced in some state which is a linear combination
of both states (66)
\begin{equation}
\left| \Lambda \left( 0\right) \right\rangle =a\left| \Lambda
_L\right\rangle +b\left| \Lambda _S\right\rangle ,\left| a\right| ^2+\left|
b\right| ^2=1. 
\end{equation}
As both ingredient states have different energy they evolve with time in a
different way, so there is some chance that after some period of time the
state $\left| \Lambda \left( 0\right) \right\rangle $ will oscillate to the
orthogonal one 
\begin{equation}
\left| \Lambda ^{^{\prime }}\left( 0\right) \right\rangle =-b^{*}\left|
\Lambda _L\right\rangle +a^{*}\left| \Lambda _S\right\rangle . 
\end{equation}
Do we have a chance to recognize both states $\left| \Lambda \left( 0\right)
\right\rangle $ and $\left| \Lambda ^{^{\prime }}\left( 0\right)
\right\rangle $? In the neutral bosons system, because of the strangeness
conservation, $K^0$ and $\overline{K}^0$interact strongly in a completely
different way and are easily distinguishable. Here we have the same particle 
$\Lambda $ with only one decay width $\Gamma _\Lambda $. In spite of that,
in principle we can distinguish $\left| \Lambda \left( 0\right)
\right\rangle $ from $\left| \Lambda ^{^{\prime }}\left( 0\right)
\right\rangle $ but in a much more sophisticated way. $\Lambda $'s in both
states will decay in the same way (mostly to $p\pi ^{-}).$ But because two
states $\left| \Lambda _{L\left( S\right) }\right\rangle $ have slightly
different momenta (in CM frame) also the angular distribution (e.g. for
protons) will be slightly different. As two states $\left| \Lambda \left(
0\right) \right\rangle $ and $\left| \Lambda ^{^{\prime }}\left( 0\right)
\right\rangle $ are various mixtures of $\left| \Lambda _{L(S)}\right\rangle
,$ the angular distribution of the protons in CM frame which come from $%
\left| \Lambda \left( 0\right) \right\rangle $ or from $\left| \Lambda
^{^{\prime }}\left( 0\right) \right\rangle $ will be different.

We can see that in principle both states (67) and (68) could be
distinguished. But do we have anything which may be distinguished? In other
words, if we have flux of $\Lambda $'s produced with kaons in the reaction
(63), will their number in the state $\left| \Lambda \left( 0\right)
\right\rangle $ or $\left| \Lambda ^{^{\prime }}\left( 0\right)
\right\rangle $ change with distance from the reaction point? For simplicity
we will present the answer to this question in the plane wave language. We
know that it is not precise, but the value of oscillation length obtained in
this way is correct. The full wave packet approach, together with particle
correlations (the EPR effect) will be presented in the next Section.

Let us assume that at $t=0,$ $\Lambda $'s are produced in the pure state $%
\left| \Lambda \left( 0\right) \right\rangle$ (in the reaction (63), the
coefficients $a=b=\frac 1{\sqrt{2}}$ , but it is more transparent to leave
them undefined). We will consider the production of $\Lambda $'s and the
decay $\Lambda \rightarrow p\pi ^{-}$ together. The amplitude for $\Lambda $
production and decay after a period of time t can be written in a form,
where two indistinguishable ways of reaching the final state are added
coherently [37]
\begin{eqnarray}
A\left( \Lambda \rightarrow p\pi ^{-}\right)& =&A\left( \Lambda \rightarrow
\Lambda _L\right) e^{-i\left( m_\Lambda -i\frac \Gamma 2\right) \tau
_L}A\left( \Lambda _L\rightarrow p\pi ^{-}\right) \nonumber \\  
&+&A\left( A\rightarrow \Lambda _S\right) e^{-i\left( m_\Lambda -i\frac \Gamma
2\right) \tau _S}A\left( \Lambda _S\rightarrow p\pi ^{-}\right) , 
\end{eqnarray}
where $A\left( \Lambda \rightarrow \Lambda _{L,S}\right) $ are amplitudes
for $\Lambda $ production in the states $\left| \Lambda _{L(S)}\right\rangle 
$, $A\left( \Lambda _{L,S}\rightarrow p\pi ^{-}\right) $ are decay
amplitudes from both states, $m_\Gamma $ and $\Gamma $ are mass and decay
width of the $\Lambda $. The different proper times which elapse in the $%
\Lambda $'s rest frames during the propagation are the crucial points in our
discussion.

If we denote 
\begin{equation}
\frac{A\left( \Lambda _S\rightarrow p\pi ^{-}\right) }{A\left( \Lambda
_L\rightarrow p\pi ^{-}\right) }=\eta _{SL}=\left| \eta _{SL}\right|
e^{i\rho _{LS}},a=\left| a\right| e^{i\varphi _a},b=\left| b\right|
e^{i\varphi _b}, 
\end{equation}
the probability for $\Lambda $ production and decay from the initial state $%
\left| \Lambda \left( 0\right) \right\rangle $ can be written 
\begin{eqnarray}
P\left( \Lambda \rightarrow p\pi ^{-}\right)& =&\left| A\left( \Lambda
\rightarrow p\pi ^{-}\right) \right| ^2=\left| A\left( \Lambda _L\rightarrow
p\pi ^{-}\right) \right| ^2   \\  
&\times&\{\left| a\right| ^2e^{-\Gamma \tau _L}+\left| b\right| ^2\left| \eta
_{SL}\right| ^2e^{-\Gamma \tau _S} \nonumber \\  
&+&2\left| ab\eta _{LS}\right| e^{-\frac 12\left( \tau _L+\tau _S\right) \Gamma
}\cos \left[ m_\Lambda \left( \tau _L-\tau _S\right) +\varphi _b+\rho
_{LS}-\varphi _a\right] \} \nonumber
\end{eqnarray}
The oscillation can possibly arise from the term $m_\Lambda \left( \tau
_L-\tau _S\right) $. If instead of $\Lambda $ we consider the production and
decay of the initial kaon a similar formula would be obtained
but with one, as we will see, crucial difference. As masses of $K_L-K_{S%
\mbox{ }}$bosons are different, the oscillation factor is equal to $m_L\tau
_L-m_S\tau _S$.

How to calculate the proper times? They are measured in different Lorentz
frames, in the rest frames of $\Lambda _L$ and $\Lambda _S$. The basic
principle of quantum mechanics - the superposition principle tells, that we
can add two states at the same time 
\begin{equation}
\left| \Psi (t)\right\rangle =\left| \varphi _1(t)\right\rangle +\left|
\varphi _2(t)\right\rangle . 
\end{equation}
In the position representation we add wave functions 
\begin{equation}
\left\langle \stackrel{\rightarrow }{x}|\Psi (t)\right\rangle =\left\langle 
\stackrel{\rightarrow }{x}|\varphi _1(t)\right\rangle +\left\langle 
\stackrel{\rightarrow }{x}|\varphi _2(t)\right\rangle 
\end{equation}
in the same position $\stackrel{\rightarrow }{x}$ and at the same time t. It
means that the proper times are not suitable variables. We have to add wave
function at the same point $\left( \stackrel{\rightarrow }{x},t\right) $ of
the same Lorentz system. It is necessary to transform $\tau _L$ and $\tau _S$
to the same common frame. The CM frame for the whole $\pi ^{-}p\rightarrow
\Lambda ^0K^{0\mbox{ }}$ process is the most convenient in this place.

For convenience, we consider only ''the one dimensional'' problem $\stackrel{%
\rightarrow }{x}=\left( x,0,0\right) $. Then Lorentz transformations between
the rest frames for $\Lambda _{L(S)}$ and the CM frame are given by

$$
\tau _L=\gamma _L\left( t-\beta _Lx\right) ,\;\xi _L=\gamma _L\left( x-\beta
_Lt\right) , 
$$

and

$$
\tau _S=\gamma _S\left( t-\beta _Sx\right) ,\;\xi _S=\gamma _S\left( x-\beta
_St\right) , 
$$
where 
\begin{equation}
\gamma _{L,S}=\frac{E_{L,S}}{m_\Lambda },\;\beta _{L,S}=\frac{p_{L,S}}{E_{L,S}}%
. 
\end{equation}
At the beginning $t=0,x=0,$ and two ''ingredients'' of the $\Lambda $
particle, $\Lambda _L$ and $\Lambda _S$ are created at $\tau _L=\tau _S=0$
and $\xi _L=\xi _S=0.$ But particles in two states $\Lambda _{L\mbox{ }}$and 
$\Lambda _{S\mbox{ }}$have different speeds and after time t they are in
different points in the CM frame (see Fig.3). 

In classical mechanics, for
point particles, it is impossible to have a situation that two particles
which were born in the same point and at the same time but moving with
different speeds would be still in the same, common points at the same time
later (Fig.3). Accordingly to our previous statement (Eq.(73)) such
particles will not interfere for any time $t>0.$ But in QM particles are
described by wave packets (in the limiting case-plane waves). We do not know
at what place the particle was born inside the wave packet and what was the
speed of it (see Fig.4). 
It is not strange that different parts of two wave
packets still interfere. Inside wave packets, energy and momentum are
distributed in agreement with QM prescriptions and it is not a surprise that
they can be not conserved [32].

As we remember, in the wave packet approach, to find the probability as a
function of position it is not necessary to assume any relation between t
and x. We simply integrate over t [36]. It is possible, however, to find
such a relation between t and x that the oscillation length, which we obtain
in this frame, will be (to the first order) the same as in a proper wave
packet approach. Such a frame was found [35]. It is the CM frame for $%
\Lambda _L$ and $\Lambda _S,$ where their momenta are opposite $p_L^{*}=-$ $%
p_S^{*}$ (see Fig.4). The velocity $\beta $ of the origin of $\Lambda
_L-\Lambda _S$ center of mass frame in the laboratory system can be easily
found from the relation 
\begin{equation}
\gamma \left( p_L-\beta E_L\right) =p_L^{*}=-p_S^{*}=-\gamma \left(
p_S-\beta E_S\right) , 
\end{equation}
so 
\begin{equation}
\beta =\frac{p_L+p_S}{E_L+E_S}=\frac{E_L-E_S}{p_L-p_S}. 
\end{equation}
Then the movement of the origin of the $\Lambda _L-\Lambda _S$ CM frame in
our laboratory system is described by the obvious relation 
\begin{equation}
x=\beta t 
\end{equation}
with $\beta $ given by Eq.(76).

Using this classical relation Eq.(77), the proper times $\tau _{L\mbox{ }}$%
and $\tau _{S\mbox{ }}($from Eq.(74)) are given by 
\begin{equation}
\begin{array}{c}
\tau _{L 
\mbox{ }}=\gamma _L\left( \frac 1\beta -\beta _L\right) x, \\  \\ 
\tau _{S\mbox{ }}=\gamma _S\left( \frac 1\beta -\beta _S\right) x 
\end{array}
\end{equation}
and taking Eqs. (74) and (77) we have [35] 
\begin{equation}
\tau _{L\mbox{ }}-\tau _{S\mbox{ }}=x\left[ \gamma _L\left( \frac 1\beta
-\beta _L\right) -\gamma _S\left( \frac 1\beta -\beta _L\right) \right] =0 
\end{equation}
and the oscillation length (Eq.(71)) is infinitely large. If we calculate,
in a analogous frame, the oscillation factor for kaons we obtain [35] 
\begin{eqnarray*}
m_L\tau _{L 
\mbox{ }}-m_S\tau _{S\mbox{ }}&=&m_L\gamma _{K_L}\left( \frac 1{\beta
_K}-\beta _{K_L}\right) -m_S\gamma _{K_S}\left( \frac 1{\beta _S}-\beta
_{K_S}\right)  \\ 
&=&\frac{m_L^2-m_S^2}{p_{K_L}+p_{K_S}}=\frac{\frac{m_L+m_S}2\left(
m_L-m_S\right) }{(p_{K_L}+p_{K_S})/2}=\frac{m\Delta m}p\equiv \frac{\Delta
m^2}{2p}, 
\end{eqnarray*}
\begin{equation}
\end{equation}
which reproduces the well known result for the oscillation frequency (Eq.(6)) 
\begin{equation}
\Delta mt=\Delta m\frac xv=\frac{m\Delta m}{mv}x=\frac{\Delta m^2}{2p}x. 
\end{equation}
The relations (79) and (80) which characterize the oscillation length are
obtained in this special Lorentz frame. In other frames, these results are
correct only to first order in $\Delta m$. Instead of checking the
dependence of the oscillation length on the Lorentz frame, in the next
Section we will present the more complete wave packet approach. Here we have
found that the oscillation length of particles recoiling against mixed
states is very large. It means that even if we consider the oscillation of
such particles $(\Lambda $ or $\mu )$ separately, without connection to
kaons or neutrinos it is impossible to observe such oscillations on any
acceptable terrestrial distance. Now we consider the oscillation of both
particles $(\Lambda $ and $K$ or $\mu $ and $\nu )$ together.

\section{Correlations for two oscillating particles, EPR effect.}

Up to now we have considered the oscillation of one particle without taking
into account possible correlations which may appear for two or more
particles in the final states from which at least one oscillates in the
traditional way. There are many such cases. Some of them, with one
oscillating particle, have been discussed above $\left( \pi ^{-}p\rightarrow
\Lambda ^0K^0\mbox{ or }\pi ^{+}\rightarrow \mu ^{+}\nu _\mu \right) .$
There are also interesting processes with two oscillating bosons e.g. $\Phi
\rightarrow K^0\overline{K}^0,\Psi \left( 3770\right) \rightarrow D^0%
\overline{D}^0$ or $e^{+}e^{-}\rightarrow \Upsilon \left( 4s\right)
\rightarrow B^0\overline{B}^0.$

Let us describe the last of them. At $t=0$ the state of two bosons is a
combination of states with definite masses 
\begin{equation}
\left| B^0\overline{B}^0\right\rangle _{t=0}=\sum_{a,b}\eta _{ab}R_{Ba}R_{%
\overline{B}b}\left| B_aB_b\right\rangle _{t=0}, 
\end{equation}
where $R_{Ba}$ and $R_{\overline{B}b}$ are elements of unitary matrix which
describes the mixing. The momentum conservation in the production process
gives altogether $\frac{n\left( n+1\right) }2$ independent momenta for all $%
n^2$ pairs $B_aB_b$. Sometimes there are additional correlations between
various mass states $B_a$ and $B_b$ in Eq.(82). If, for example, the $B^0%
\overline{B}^0$pairs are produced by the $\Upsilon (4s)$ decay then the
state $\left| B^0\overline{B}^0\right\rangle $ must be totally antisymmetric
[38,39] (since $\Upsilon (4s)$ has intrinsic spin $s=1$ but B mesons are
spinless, the B pair is in a p wave). The factors $\eta _{ab}$ in Eq.(71)
are responsible for such correlations (see for details Ref.[36]).

Each state $\left| B_{a\left( b\right) }\right\rangle $ is described by a
wave packet which in the momentum representation is given by 
\begin{equation}
\left\langle a,\stackrel{\rightarrow }{p}|B_a\right\rangle _{t=0}=\int
d^3p\Psi _a\left( \stackrel{\rightarrow }{p},\stackrel{\rightarrow }{p}_a%
,\sigma _p\right) \left| a,\stackrel{\rightarrow }{p}\right\rangle , 
\end{equation}
where $\Psi _a\left( \stackrel{\rightarrow }{p},\stackrel{\rightarrow }{p}_a%
,\sigma _p\right) $ for simplicity is taken as the Gauss function 
\begin{equation}
\Psi _a\left( \stackrel{\rightarrow }{p},\stackrel{\rightarrow }{p}_a,\sigma
_p\right) =\frac 1{\left[ \sqrt{2\pi }\sigma _p\right] ^{3/2}}e^{-\frac{%
\left( \stackrel{\rightarrow }{p}-\stackrel{\rightarrow }{p}_a\right) ^2}{%
\scriptstyle 4\sigma _p^2}}, 
\end{equation}
with $\stackrel{\rightarrow }{p_a}$ - the average momentum and $\sigma _p$ -
the width of the distribution. After time t (taking into account the
particle decay) the state $\left| B_a\right\rangle $ will evolve into 
\begin{eqnarray}
\left| B_a\right\rangle _{t=0} & \rightarrow & \left| B_a\left( t\right)
\right\rangle \cong e^{-iHt}\left| B_a\right\rangle _{t=0} \nonumber \\  
&=& e^{-\frac t{2\tau _a}}\int d^3p\Psi _a\left( \stackrel{\rightarrow }{p},%
\stackrel{\rightarrow }{p}_a,\sigma _p\right) e^{-iE_a ( \stackrel{%
\rightarrow }{p}) t}\left| a,\stackrel{\rightarrow }{p}\right\rangle , 
\end{eqnarray}
where $\tau _a=\frac 1{\Gamma _a}\left( \frac{E_a\left( \stackrel{\rightarrow 
}{p}_a\right) }{m_a}\right) $ is the lifetime of the $"a"$ particle in a
chosen Lorentz frame.

The states $\left| B_a\left( t\right) \right\rangle $ in the position
representation will be given by 
\begin{equation}
\left| B_a\left( t\right) \right\rangle =e^{-\frac t{2\tau _a}}\int d^3x\Psi
_a\left( \stackrel{\rightarrow }{x},t;\stackrel{\rightarrow }{v_a,}\sigma
_{BP}\right) \left| a,\stackrel{\rightarrow }{x}\right\rangle , 
\end{equation}
where $\Psi _a\left( \stackrel{\rightarrow }{x},t;\stackrel{\rightarrow }{%
v_a,}\sigma _{BP}\right) $ is the Fourier transform of the momentum
distribution (Eq.(84)) and is given by Eq.(25). Let us assume that two
detectors are placed at points $\stackrel{\rightarrow }{L_1}$ and $\stackrel{%
\rightarrow }{L_2}$. The detectors will measure the particles with beauty
''1'' and ''2''  ($1,2=B^0,\overline{B}^0 $, respectively).

The states of the B mesons measured by the two detectors are defined by 
\begin{equation}
\left| B_1\left( \stackrel{\rightarrow }{L_1}\right) \right\rangle
=\sum_cR_{1c}\int d^3x_1\Psi _c\left( \stackrel{\rightarrow }{x_1}-\stackrel{%
\rightarrow }{L_1},0;\stackrel{\rightarrow }{v_c},\sigma _{1D}\right) \left|
c,\stackrel{\rightarrow }{x_1}\right\rangle , 
\end{equation}
and 
\begin{equation}
\left| B_2\left( \stackrel{\rightarrow }{L_2}\right) \right\rangle
=\sum_dR_{2d}\int d^3x_2\Psi _d\left( \stackrel{\rightarrow }{x_2}-\stackrel{%
\rightarrow }{L_2},0;\stackrel{\rightarrow }{v_d},\sigma _{2D}\right) \left|
d,\stackrel{\rightarrow }{x_2}\right\rangle . 
\end{equation}
Notations in Eqs. (87) and (88) are similar to previously presented in Eq.(30).

We can find the amplitude for two oscillating particles in the same way as
before (Chapter 3). Then the amplitude of the process where two particles $B^0$
and $\overline{B}^0$ produced at $t=0$ at point $\stackrel{\rightarrow }{x}%
=0 $ are detected as particles with beauty ''1'' (''2'') at point $L_1 
(L_2)$ at $%
t=t_B\left( t_{\overline{B}}\right) $, is the following 
\begin{eqnarray}
&&A_{B^0\rightarrow B_1, 
\overline{B}^0\rightarrow B_2}(\stackrel{\rightarrow }{L_1},t_B;\stackrel{%
\rightarrow }{L_2},t_{\overline{B}})= 
\left\langle B_1\left( 
\stackrel{\rightarrow }{L_1}\right) ,B_2\left( \stackrel{\rightarrow }{L_2}%
\right) \mid B^0\left( t_B\right) ,\overline{B}^0\left( t_{\overline{B}%
}\right) \right\rangle \nonumber  \\ 
&=& N\sum_{a,b}R_{1a}^{*}R_{Ba}R_{2b}^{*}R_{Bb}\eta
_{ab}e^{-\frac 12\left( 
\frac{t_B}{\tau _a}+\frac{t_{\overline{B}}}{\tau _b}\right) } \nonumber  \\ 
&\times& e^{-i\left( E_at_B-\stackrel{\rightarrow }{p_a}\stackrel{\rightarrow }{L_1}%
\right) -i\left( E_bt_{\overline{B}}-\stackrel{\rightarrow }{p_b}\stackrel{%
\rightarrow }{L_2}\right) }e^{-\frac{\left( \stackrel{\rightarrow }{L_1}-%
\stackrel{\rightarrow }{v_a}t_B\right) ^2}{4\sigma _B}-\frac{\left( 
\stackrel{\rightarrow }{L_2}-\stackrel{\rightarrow }{v_b}t_{\overline{B}%
}\right) ^2}{4\sigma _{\overline{B}}}}, \nonumber
\end{eqnarray}
\begin{equation}
\end{equation}
with the normalization factor 
\begin{equation}
N=\left[ \frac{4\sigma _{1D}\sigma _{2D}\sigma _{BP}\sigma _{\overline{B}P}}{%
\sigma _B\sigma _{\overline{B}}}\right] ^{1/2}, 
\end{equation}
and the effective total widths 
\begin{equation}
\sigma _B=\sqrt{\sigma _{1D}^2+\sigma _{BP}^2},\;\;\sigma _{\overline{B}}=\sqrt{%
\sigma _{2D}^2+\sigma _{\overline{B}P}^2}. 
\end{equation}
The amplitude (89) can be used in various situations. If we apply the
formula (89) to the description of the EPR effect in the $\Upsilon \left(
4s\right) \rightarrow BB$ decay (Refs [38,39]), the collapse of the BB wave
function is included in a natural way. Our approach is an alternative to the
amplitude description (Refs [38,39]) and, in case of particle mixing, takes
into account the EPR correlations in a much more transparent way (see Ref.[36] for detail).

The formula (89) can also be used for the $\mu \nu _\mu $ pair
''oscillation'' from the $\pi \rightarrow \mu \nu _\mu $ decay. In this case
only one particle, the neutrino, mixes. Then, in the application of Eq.(89)
to our present purpose, we have to take one diagonal mixing matrix (e.g. $%
R_{2b}=\delta _{2b},R_{\overline{B}b}=\delta _{\overline{B}b}$) and $\eta
_{ab}=\delta _{ab}$. Usually neutrinos are considered as stable or very long
living particles. Let us also assume that the ''oscillation'' of the muons
is measured on a distance much shorter than their decay length. Then both
factors in Eq.(89), which are responsible for particle decay, may be
neglected. In such circumstances, the probability that neutrinos produced as 
$\nu _\mu =\alpha $ type together with muons at $\stackrel{\rightarrow }{x}%
=0 $ are observed as a $\beta $-type neutrino at distance $L_\nu$ 
and the muons at distance $L_\mu,$ after integrating over times is given by (for details see 
Ref.[36]), 
\begin{equation}
\begin{array}{c}
P_{\beta \alpha }\left( L_\mu ,L_\nu \right) =\left[ \sum\limits_a 
\frac{U_{\alpha a}}{v_{\mu a}v_{\nu a}}\right] ^{-1}\sum\limits_{ab}\sqrt{\frac
4{\left( v_{\mu a}^2+v_{\mu b}^2\right) \left( v_{\nu a}^2+v_{\nu
b}^2\right) }}\times U_{\beta b}U_{\alpha b}^{*}U_{\beta a}^{*}U_{\alpha
a}\times \\  \\ 
e^{-2\pi i\left( \frac{L_\mu }{L_{ab}^{\mu osc}}\right) }e^{-\left( \frac{%
L_\mu }{L_{ab}^{\mu coh}}\right) ^2}e^{-\left( \frac{\sigma _{\mu x}}{%
L_{ab}^{\mu osc}}\right) ^2N_{ab}^\mu }\times e^{-2\pi i\left( \frac{L_\nu 
}{L_{ab}^{\nu osc}}\right) }e^{-\left( \frac{L_\nu }{L_{ab}^{\nu coh}}%
\right) ^2}e^{-\left( \frac{\sigma _{\nu x}}{L_{ab}^{\nu osc}}\right)
^2N_{ab}^\nu }. 
\end{array}
\end{equation}
The oscillation $\left( L_{ab}^{\mu osc}\right) $and coherence $\left(
L_{ab}^{\mu coh}\right) $lengths, and the factor $N_{ab}^\mu $ for muons are
(for neutrinos the appropriate expressions are similar) 
\begin{equation}
L_{ab}^{\mu osc}=2\pi \left[ \left( E_{\mu a}-E_{\mu b}\right) \frac{v_{\mu
a}+v_{\mu b}}{v_{\mu a}^2+v_{\mu b}^2}-\left( p_{\mu a}-p_{\mu b}\right)
\right] ^{-1}, 
\end{equation}
\begin{equation}
L_{ab}^{\mu coh}=2\sigma _{\mu x}\left( \frac{v_{\mu a}^2+v_{\mu b}^2}{%
\left( v_{\mu a}-v_{\mu b}\right) ^2}\right) ^{1/2}, 
\end{equation}
and 
\begin{equation}
N_{ab}^\mu =4\pi ^2\frac{\left( E_{\mu a}-E_{\mu b}\right) ^2\left( v_{\mu
a}^2+v_{\mu b}^2\right) }{\left[ \left( E_{\mu a}-E_{\mu b}\right) \left(
v_{\mu a}+v_{\mu b}\right) -\left( p_{\mu a}-p_{\mu b}\right) \left( v_{\mu
a}^2+v_{\mu b}^2\right) \right] ^2}, 
\end{equation}
where $E_{\mu a\left( b\right) },p_{\mu a\left( b\right) }$ and $v_{\mu
a\left( b\right) }$ are energy, momentum and velocity of the muon associated
with the neutrino a(b).

First of all, we can see from Eq.(92) that the muon oscillation disappears
if we do not measure separately the neutrinos with flavour $\beta $. In such
case the probability given by amplitude (92) is constant in $L_\mu $ and $%
L_\nu $ and can be normalized 
\begin{equation}
\sum_\beta P_{\beta \alpha }\left( L_\mu ,L_\nu \right) =1. 
\end{equation}
But even if we measure the $\beta $-type neutrino, the muon oscillation will
not be seen. We can prove this statement because we know precisely the muon
oscillation length (Eq.(93)).

If we denote the difference between the masses of two neutrinos a and b, as 
\begin{equation}
m_a-m_b=\Delta m_{ab}, 
\end{equation}
the inverse of the oscillation length may be decomposed in powers of $\Delta
m_{ab}$. For muons, the term proportional to the first power of $\Delta
m_{ab}$ vanishes, and 
\begin{equation}
\left( 2\pi \right) \left( L_{ab}^{\mu osc}\right) ^{-1}=-2\zeta ^2\left(
1-v_a^2\right) p_a^{-1}\left( \Delta m_{ab}\right) ^2+..., 
\end{equation}
where $\zeta $ is the factor in the decomposition 
\begin{equation}
p_b=p_a\left( 1+\zeta \left( \frac{\Delta m_{ab}}{p_a}\right) +\rho \left( 
\frac{\Delta m_{ab}}{p_a}\right) ^2+...\right) , 
\end{equation}
while for neutrinos there is 
\begin{equation}
\left( 2\pi \right) \left( L_{ab}^{\nu osc}\right) ^{-1}=\frac{\Delta
m_{ab}^2}{2p_a}+.... 
\end{equation}
and we reconstruct the previous formula for mixing particle oscillation
length Eqs.(12), (38).

From Eq.(98) it follows that $L_{ab}^{\mu osc}$ is very large and for the
acceptable neutrino mass difference the muon oscillation length is much
bigger than its decay length 
\begin{equation}
L_{ab}^{\mu osc}\gg c\tau _\mu \approx 660 \mbox{\rm m}. 
\end{equation}
We can see that even if the neutrino and the muon are both measured, the
oscillation of muon will not be observed. Taking into account opinions
presented in the latest exchange of views [31-35] we agree with the
statement that, in practice, oscillations of $\mu $ or $\Lambda
^{0\mbox{ }}$%
particle are impossible to observe.

\section{Summary}

Let us briefly summarize the main results of this review paper

{\bf (i)} Many conceptual difficulties arise in the plane wave approach to
the particle oscillation problem. This approach gives us the shortest way to
get the correct expression for the oscillation length, but it fails if we
try to describe other aspects of the particle oscillation.

{\bf (ii)} In real oscillation experiments neither energy nor momentum are
the same for all eigenmass particles.

{\bf (iii)} The wave packet approach

- gives the proper oscillation length $L_{ab}^{osc},$

- introduces the concept of the coherence length $L_{ab}^{coh},$ such that
for distances greater than $L_{ab}^{coh}$ the particle oscillation
disappears,

- in the proper way, takes into account the fact that to observe
oscillation, the sizes of particles source and detector must be much smaller
than the oscillation length,

- gives a possibility to understand in a simple way the phenomenon, that a
precise measurement of the detected particles momenta may restore the
coherence between various eigenmass states and, as a consequence, the
oscillation between particles,

- temporal and spatial distributions in the wave packet are correlated by
the requirement, that particles are on mass shell. Independent distributions
for time and space give a wave packet which does not satisfy the free
particle wave equation.

{\bf (iv)} A problem arises with the proper definition of the Fock space for
flavour states. As a consequence, the defined probability currents for such
states are not conserved. Then the calculated flavour changing probability
is correct only to the first power of the mass difference $\Delta m.$

{\bf (v)} The most adequate approach to particle oscillation is given by
quantum field theory. It can be seen especially for neutrinos, which are
''neither prepared nor observed'', and only propagate between sources and
detectors. In this approach

- production and detection processes are fully taken into account as in real
experiments,

- physical quantities are expressed in terms of measured quantities, like
momenta and energies of hadrons or charged leptons in the neutrino creation
and detection processes,

- flavour states, which are not well defined, are not necessary to describe
the oscillation process. Only the mass eigenstates and the elements of the
mixing matrices may be used,

- it is clear in which circumstances the production and detection amplitudes
can be factorized out and the separate oscillation probability be defined,

- the oscillation of non- relativistic particles can be described in the
proper way.

{\bf (vi)} We have specified the meaning of oscillation of particles which
recoil against mixed states (as $\Lambda $ in the process $\pi
^{-}p\rightarrow \Lambda ^0K^0$ or $\mu $ in the decay $\pi ^{+}\rightarrow
\mu ^{+}\nu _\mu $). Even if we consider oscillations of such particles
separately, without connection to kaons or neutrinos, it is impossible to
observe these oscillations on acceptable terrestrial distances.

{\bf (vii)} We have found the general amplitudes which describe the
oscillation of two particles in the final states. These amplitudes can be
applied to

- description of the EPR correlations in decays like $\Upsilon
(4s)\rightarrow BB$ or $\Phi \rightarrow KK$, including the mysterious
collapse of the wave function in a natural way and giving the possibility to
discuss the relativistic EPR correlations on distances longer than coherence
lengths.

- description of two particles oscillation from which only one has
indeterminate mass like $\Lambda K^0$ or $\mu \nu _\mu $ . Oscillations of
particles with known mass (eq. $\Lambda $ or $\mu $) can be defined only if,
in the same time, flavours of the unknown mass particles are measured ($K^0$
or $\nu _\mu $). In this case, however, the oscillation length of particles
with determinate mass is very large, much larger than the particle decay
length, which makes it impossible to observe their oscillation in practice.

I thank J. S\l adkowski for valuable remarks and M. Czakon and J. Gluza 
for reading the text.

\begin{figure}
\center
\psfig{figure=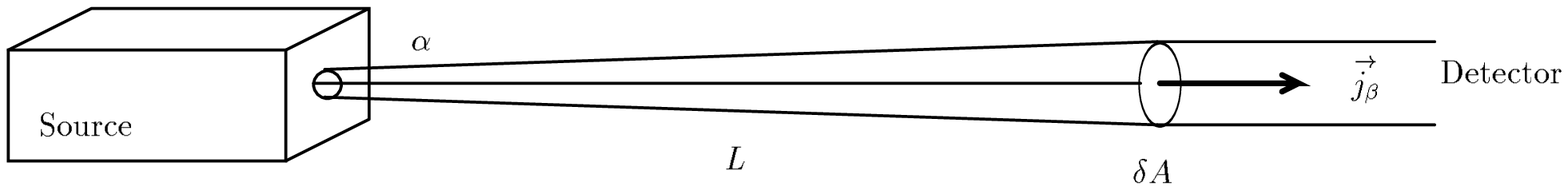,height=.7 in}
\caption{Particles with the flavour $\alpha$ are produced in the
"Source". After travelling the distance L they are detected as particles
with flavour $\beta$. The probability of such detection is given by the
probability current integral over the detector active surface $(\delta A)$
and over the time of measurements.}
\end{figure}

\begin{figure}
\center
\psfig{figure=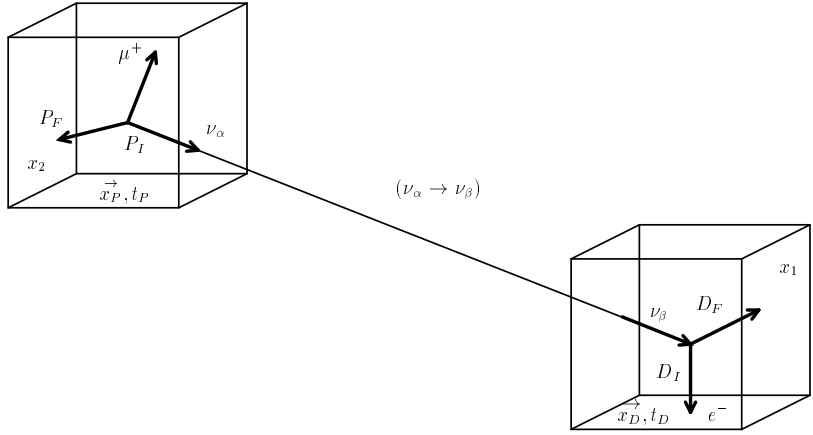,height=2.5 in}
\caption{ In the field theory approach to the neutrino oscillation only
the source $(P_I,P_F,\mu^+)$ and detector $(D_I,D_F,e^-)$ particles are
measured. Between the production $(\vec{x}_P,t_P)$ and detection
$(\vec{x}_D,t_D)$ points neutrino propagates as virtual particle.}
\end{figure}

\begin{figure}
\center
\psfig{figure=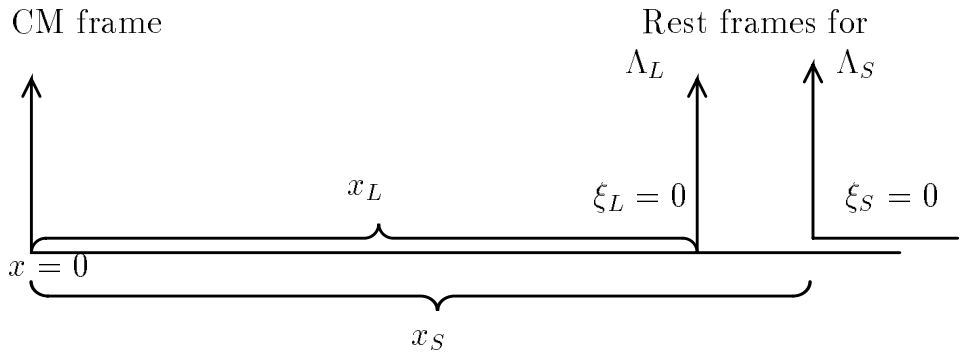,height=2.5 in}
\caption{ Relation between CM frame for the production process $\pi^- p
\rightarrow K^0 \Lambda^0$ and the two rest frames for the $\Lambda_L$ and
$\Lambda_S$ which move with different speeds. In classical mechanics for
$t>0$ the $\Lambda_L$ and $\Lambda_S$ are in different points. }
\end{figure}

\begin{figure}
\center
\psfig{figure=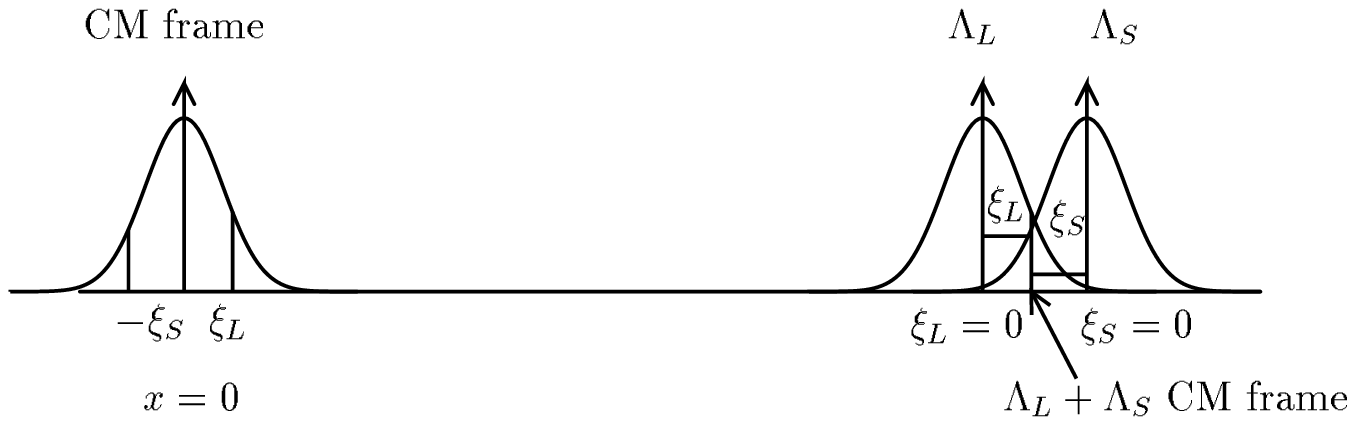,height=2.5 in}
\caption{ The same as in Fig.3 but in quantum mechanics where the
$\Lambda_L$ and $\Lambda_S$ particles are described by the wave packets.
Even if the centres of the wave packets  at
the same time are in the different points (like for classical particle in
Fig.3) the two states for $\Lambda_L$ and $\Lambda_S$ may interfere. The
interference will be possible if the two wave packets overlap.} 
\end{figure}


\begin{thebibliography}{99}
\bibitem{} M. Gell Mann, A. Pais, Phys. Rev. 97 (1995) 1387.
\bibitem{} J. H. Christensen, J.W. Cronin, V.L. Fitch, R. Turlay, Phys. Rev.
Lett. 13 (1964) 138.
\bibitem{} B. Pontecorvo, Zh. Exp. Fiz. 33 (1957) 549.
\bibitem{} Z. Maki, M. Nakagawa, S. Sakata, Prog. Theor. Phys. 28 (1962) 870.
\bibitem{} V. Gribov, B. Pontecorvo, Phys. Lett. B28 (1969) 493;
J.N. Bahcall, S. Frautschi, Phys. Lett. B29 (1969) 623;
S. Eliezer, D.A. Ross, Phys. Rev. D10 (1974) 3088;
S. Eliezer, A.R. Swift, Nucl. Phys. B105 (1976) 45;
S.M. Bilenky, B. Pontecorvo, Phys. Lett. B61 (1976) 248;
H. Fritzsch, P. Minkowski, Phys. Lett. B62 (1976) 72;
A.K. Mann, H. Primakoff, Phys. Rev. D15 (1977) 655.
\bibitem{} R.H. Good et al., Phys. Rev. 124 (1961) 1223.
\bibitem{} C. Geweniger et al., Phys. Lett. B48 (1974) 487;
for the recent test see, for example
R. Adler et al. (CPLEAR Collaboration), Phys. Lett. B363 (1995) 237;
Phys. Lett. B363 (1995) 243;
D. Buskulic et al. (ALEPH Collaboration), Z. Phys. C75 (1997) 397.
\bibitem{} B. Pontecorvo, Sov. Phys. JETP 26 (1968) 984;
L. Wolfenstein, Phys. Rev. D17 (1978) 2369; Phys. Rev. D20 (1979)
2634; S.P. Mikheyev, A. Yu. Smirnov,Yad. Fiz. 42 (1985) 1441;
Il Nuovo Cimento C9 (1986) 17;
see also: J.N. Bahcall, ''Neutrino Astrophysics'' ed. Cambridge
University Press, Cambridge, 1989;
for the last data see: Superkamiokande Collaboration, hep-ex/9805021.
\bibitem{} C. Athanassopoulos et al. (LSND), Phys. Rev. Lett. 75 (1995) 2650;
ibid. 77 (1996) 3082; nucl-ex/9706006;
for oscillation of atmospheric neutrinos see:
Superkamiokande Collaboration, hep-ex/9803006,hep-ex/9805006;
hep-ex/9807003.
\bibitem{} V. Bargmann, Annals of Math.59(1954)1; see also: A. Galindo,P. Pascual,
''Quantum Mechanics'', Springer Verlag, 1990, p. 288.
\bibitem{} See for example:
E.D. Commins, P.H. Bucksbaum, ''Weak Interaction of Leptons and
Quarks'', Cambridge University Press, Cambridge, 1983, p. 247;
W.E. Burcham and M. Jobes, ''Nuclear and Particle Physics'', Longmans,
Harlow, UK, 1995.
\bibitem{} See for example:
S.M. Bilenky, S.T. Petcov, Rev. Mod. Phys. 59 (1987) 671;
B. Kayser, F. Gibrat-Debu, F. Perrier, ''The Physics of Massive
Neutrinos'',World Scientific, Singapore, 1988, p.10;
R.N. Mohapatra, P.B. Pal, ''Massive Neutrinos in Physics and
Astrophysics'',World Scientific, 1991, p. 156;
T.P. Cheng and L.F. Li, ''Gauge Theory of Elementary Particle Physics'',
Clarendon Press, Oxford, 1984, p. 410.
\bibitem{} B. Kayser, Phys. Rev. D24 (1981) 110.
\bibitem{} R.G. Winter, Lett. Nuovo Cim. 30 (1981) 101;
F.Boehm, P. Vogel, ''Physics of Massive Neutrinos'', Cambridge Univ.
Press,1992, p. 92;
T. Goldman, LA-UR-96-1349,hep-ph/9604357.
\bibitem{} K. Grotz, H.V. Klapdor, The Weak Interaction in Nuclear,
Particle and Astrophysics'', Adam Higler, Bristol, 1990, p. 296.
\bibitem{} H. J. Lipkin, Phys. Lett. B348 (1995) 604.
\bibitem{} Y. Grossman and H.J. Lipkin, Phys. Rev. D55 (1997) 2760.
\bibitem{} B. Ancochea, A. Bramon, R. Munoz-Tapia, M. Nowakowski,
Phys. Lett. B389 (1996) 149.
\bibitem{} J. Rich, Phys. Rev. D48 (1993) 4318.
\bibitem{} C. Giunti, C.W. Kim, J.A. Lee, U.W. Lee, Phys. Rev. D48
(1993) 4310.
\bibitem{} C. Giunti, C.W. Kim, U.W. Lee, Phys. Rev. D44 (1991) 3635;
C.W. Kim and A. Pevsner, ''Neutrinos in Physics and Astrophysics'',
Contemporary Concepts in Physics, vol. 8 ed. by H. Feshbach
(Harwood Academic Chur), Switzerland, 1993.
\bibitem{} C. Giunti, C.W. Kim, Phys. Rev. D58 (1998) 017301;
hep-ph/9711363.
\bibitem{} S. Nussinov, Phys. Lett. B63 (1976) 201.
\bibitem{} K. Kiers, S. Nussinov, Weiss, Phys. Rev. D53 (1996) 537;
K.Kiers, N. Weiss, Phys. Rev. D57 (1998) 3091; hep-ph/9710289.
\bibitem{} S. Mohanty, hep-ph/9702428; hep-ph/9706328; hep-ph/9710284.
\bibitem{} M. Blasone, G. Vitiello, Ann.Phys. (N.Y.) 244 (1995) 283;
E. Alfinito, M. Blasone, A. Iorio, G. Vitiello, Acta Phys. Pol. 27B (1996)
1493; hep-ph/9510213;
M. Blasone, G. Vitiello, Annals Phys. 244 (1995) 283, Erratum- ibid. 249
(1996) 249; hep-ph/9501263;
M. Blasone, P.A. Henning, G. Vitiello, in Proceedings of ''Results
and Perspectives in Particle Physics'', La Thuile, Aosta Valley, March
1996;
E. Sassaroli, hep-ph/9609476; hep-ph/9805480; M. Blasone, hep-ph/9810329; F.
Fujii, Ch. Haba, T. Yabuki, hep-ph/9807266.
\bibitem{} C. Giunti, C.W. Kim, U.W. Lee, Phys. Rev. D45 (1992) 2414.
\bibitem{} W. Grimus, P. Stockinger, Phys. Rev. D54 (1996) 3414.
\bibitem{} Yu. V. Shtanov, Phys. Rev. D57 (1998) 4418; hep-ph/9706378.
\bibitem{} C. Giunti, C.W. Kim, U.W.\ Lee, Phys. Lett. B421 (1998) 237;
hep-ph/9709494.
\bibitem{} Y.N. Srivastava, A. Widom, E. Sassaroli, Phys. Lett. B344 (1995)
436; hep-ph/9509261; Z. Phys. C66 (1995) 601.
\bibitem{} Y.N. Srivastava, A. Widom, hep-ph/9511294;
hep-ph/9605399; hep-ph/9612290; hep-ph/9707268.
\bibitem{} J. Lowe, B. Bassalleck, H. Burkhardt, A. Rusek, G.J. Stephenson Jr,
T. Goldman, Phys. Rev. B384 (1996) 288.
\bibitem{} A.D. Dolgov, A. Yu. Morozov, L.B. Okun, M.G. Schepkin,
Nucl. Phys. B502 (1997) 3.
\bibitem{} H. Burkhardt, J. Lowe, G.J. Stephenson Jr, T. Goldman,
hep-ph/9803365.
\bibitem{} M. Marganska, M. Zralek, in preparation.
\bibitem{} B. Kayser, L. Stodolsky, Phys. Lett. B359 (1995) 359.
\bibitem{} B. Kayser, in ''Proceedings of the Mariond Workshop on
Electroweak Interaction and Unified Theories'', Les Ares, France, March
1995; hep-ph/9509386.
\bibitem{} B. Kayser, in ''Proceedings of the 28th Conference on High
Energy Physics'', Warsaw, July 1996; hep-ph/9702327.
\end{thebibliography}
\end{document}